\documentclass[aps,showpacs,preprintnumbers,amsmath,amssymb,nofootinbib]{revtex4}
\usepackage{graphicx}

\begin{document}

\preprint{gr-qc/0404104}

\title{Can the galactic rotation curves be explained in brane world
models?}

\author{M. K. Mak}
\email{mkmak@vtc.edu.hk} \affiliation{Department of Physics, The
University of Hong Kong, Pokfulam Road, Hong Kong}

\author{T. Harko}
\email{harko@hkucc.hku.hk} \affiliation{ Department of Physics,
The University of Hong Kong, Pokfulam Road, Hong Kong}

\date{April 26, 2004}


\begin{abstract}

We consider solutions with conformal symmetry of the static,
spherically symmetric gravitational field equations in the vacuum
in the brane world scenario. By assuming that the vector field
generating the symmetry is non-static, the general solution of the
field equations on the brane can be obtained in an exact
parametric form, with the conformal factor taken as parameter. As
a physical application of the obtained solutions we consider the
behavior of the angular velocity of a test particle moving in a
stable circular orbit. In this case the tangential velocity can be
expressed as a function of the conformal factor and some
integration constants only. For a specific range of the
integration constants, the tangential velocity of the test
particle tends, in the limit of large radial distances, to a
constant value. This behavior is specific to the galactic rotation
curves, and is explained usually by invoking the hypothesis of the
dark matter. The limiting value of the angular velocity of the
test particle can be obtained as a function of the baryonic mass
and radius of the galaxy. The behavior of the dark radiation and
dark pressure terms is also considered in detail, and it is shown
that they can be expressed in terms of the rotational velocity of
a test particle. Hence all the predictions of the present model
can be tested observationally. Therefore the existence of the
non-local effects, generated by the free gravitational field of
the bulk in a conformally symmetric brane, may provide an
explanation for the dynamics of the neutral hydrogen clouds at
large distances from the galactic center.

\end{abstract}


\maketitle

\section{Introduction}

Einstein's theory of general relativity, and some of its
generalizations, proved to be in excellent agreement with the
observational or experimental results in the solar system, binary
star systems or laboratory \cite{Wi93}. However, it has long been
known that Newtonian or general relativistic mechanics applied to
the visible matter in galaxies and clusters does not correctly
describe the dynamics of those systems. The rotation curves of
spiral galaxies \cite{Bi87} are one of the best evidences showing
the problems Newtonian mechanics and/or standard general
relativity has to face on the galactic/intergalactic scale. In
these galaxies neutral hydrogen clouds are observed at large
distances from the center, much beyond the extent of the luminous
matter. Assuming a non-relativistic Doppler effect and emission
from stable circular orbits in a Newtonian gravitational field,
the frequency shifts in the $21$ cm line hydrogen emission lines
allows the measurement of the velocity of the clouds. Since the
clouds move in circular orbits with velocity $v_{tg}(r)$, the
orbits are maintained by the balance between the centrifugal
acceleration $v_{tg}^2/r$ and the gravitational attraction force
$GM(r)/r^2$ of the total mass $M(r)$ contained within the orbit.
This allows the expression of the mass profile of the galaxy in
the form $M(r)=rv_{tg}^2/G$.

Observations show that the rotational velocities increase near the
center of the galaxy and then remain nearly constant at a value of
$v_{tg\infty }\sim 200$ km/s \cite{Bi87}. This leads to a mass
profile $M(r)=rv_{tg\infty }^2/G$. Consequently, the mass within a
distance $r$ from the center of the galaxy increases linearly with
$r$, even at large distances where very little luminous matter can
be detected. This behavior of the galactic rotation curves is
explained by postulating the existence of some dark (invisible)
matter, distributed in a spherical halo around the galaxies. The
dark matter is assumed to be a cold, pressureless medium. There
are many possible candidates for dark matter, the most popular
ones being the weekly interacting massive particles (WIMP). Their
interaction cross section with normal baryonic matter, while
extremely small, are expected to be non-zero and we may expect to
detect them directly. It has also been suggested that the dark
matter in the Universe might be composed of superheavy particles,
with mass $\geq 10^{10}$ GeV. But observational results show the
dark matter can be composed of superheavy particles only if these
interact weakly with normal matter or if their mass is above
$10^{15}$ GeV \cite{AlBa03}.

From a general relativistic point of view the space-time
geometries associated with dark matter halos were considered in
\cite{Ma03}, where several properties of this space-time and the
characteristics of the possible energy-momentum tensors which
could produce such geometries have been discussed. The form of the
galactic potentials can be obtained, within a general relativistic
framework, from the observed rotation curves, without specific
reference to any metric theory of gravity. Given the potential,
the gravitational mass can be determined by way of an anisotropy
function of the static, spherically symmetric gravitational
galactic field \cite{La03}. The possibility that dark matter has a
substantial amounts of pressure, comparable in magnitude to the
energy density has been investigated in \cite{BhKa03}. Galaxy
halos models, consistent with observations of flat rotation
curves, are possible for a variety of equations of state with
anisotropic pressures.

However, despite more than 20 years of intense experimental and
observational effort, up to now no {\it non-gravitational}
evidence for dark matter has ever been found: no direct evidence
of it and no annihilation radiation from it. Moreover, accelerator
and reactor experiments do not support the physics (beyond the
standard model) on which the dark matter hypothesis is based.

Therefore, it seems that the possibility that Einstein's (and the
Newtonian) gravity breaks down at the scale of galaxies cannot be
excluded {\it a priori}. Several theoretical models, based on a
modification of Newton's law or of general relativity, have been
proposed to explain the behavior of the galactic rotation curves.
A modified gravitational potential of the form $\phi =-GM\left[
1+\alpha \exp \left( -r/r_{0}\right) \right] /\left( 1+\alpha
\right) r$, with $\alpha =-0.9$ and $r_{0}\approx 30$ kpc can
explain flat rotational curves for most of the galaxies
\cite{Sa84}.

In an other model, called MOND, and proposed by Milgrom \cite{Mi},
the Poisson equation for the gravitational potential $\nabla
^{2}\phi =4\pi G\rho $ is replaced by an equation of the form
$\nabla \left[ \mu \left( x\right) \left( \left| \nabla \phi
\right| /a_{0}\right) \right] =4\pi G\rho $, where $a_{0}$ is a
fixed constant and $\mu \left( x\right) $ a function satisfying
the conditions $\mu \left( x\right) =x$ for $x<<1$ and $\mu \left(
x\right) =1$ for $x>>1$. The force law, giving the acceleration
$a$ of a test particle  becomes $a=a_{N}$ for $a_{N}>>a_{0}$ and $a=\sqrt{%
a_{N}a_{0}}$ for $a_{N}<<a_{0},$where $a_{N}$ is the usual
Newtonian acceleration. The rotation curves of the galaxies are
predicted to be flat, and they can be calculated once the
distribution of the baryonic matter is known. Alternative
theoretical models to explain the galactic rotation curves have
been elaborated recently by Mannheim \cite{Ma93} and Moffat and
Sokolov \cite{Mo96}.

A general analysis of the possibility of an alternative gravity
theory explaining the dynamics of galactic systems without dark
matter was performed by Zhytnikov and Nester \cite{Ne94}. From
very general assumptions about the structure of a relativistic
gravity theory (the theory is metric, and invariant under general
coordinates transformation, has a good linear approximation, it
does not possess any unusual gauge freedom and it is not a higher
derivative gravity) a general expression for the metric to order
$(v/c)^2$ has been derived. This allows to compare the predictions
of the theory with various experimental data: the Newtonian limit,
light deflection and retardation, rotation of galaxies and
gravitational lensing. The general conclusion of this study is
that the possibility for any gravity theory to explain the
behavior of galaxies without dark matter is rather improbable.

The idea that our four-dimensional Universe might be a
four-dimensional space-time, embedded in a higher dimensional
space-time, had been proposed and studied, from both mathematical
and physical points of view, for a long time (for a full account
of the existing results on the subject and on the early references
see \cite{PaTa01}). The embedding approach to gravity has its
origins in the book by Eisenhart \cite{Ei49}. The mathematical
problems of the embeddings in higher dimensional space-times, with
applications to general relativity have been discussed in detail
as early as 1965 \cite{NeRo65}. By using a more physical approach
Akama \cite{Ak83} and Rubakov and Shaposhnikov \cite{RuSh83} have
suggested that we may live on a domain wall in a higher
dimensional space. Earlier references to these topics can also be
found in Bandos and Kummer \cite{BaKu99}. In this paper a
generalization of the embedding approach for $d$-dimensional
gravity based upon $p$-brane theories is considered.

Recently,  due to the proposal by Randall and Sundrum \cite{RS99a}
that our four-dimensional space-time is a three-brane, embedded in
a five-dimensional space-time (the bulk), the idea of the
embedding of our Universe in a higher dimensional space had
attracted again a considerable interest. According to the
brane-world scenario, the physical fields (electromagnetic,
Yang-Mills etc.) in our four-dimensional Universe are confined to
the three brane. These fields are assumed to arise as fluctuations
of branes in string theories. Only gravity can freely propagate in
both the brane and bulk space-times, with the gravitational
self-couplings not significantly modified. This model originated
from the study of a single $3$-brane embedded in five dimensions,
with the $5D$ metric given by $ds^{2}=e^{-f(y)}\eta _{\mu \nu
}dx^{\mu }dx^{\nu }+dy^{2}$, which, due to the appearance of the
warp factor, could produce a large hierarchy between the scale of
particle physics and gravity. Even if the fifth dimension is
uncompactified, standard $4D$ gravity is reproduced on the brane.
Hence this model allows the presence of large, or even infinite
non-compact extra dimensions. Our brane is identified to a domain
wall in a $5$-dimensional anti-de Sitter space-time. For a review
of dynamics and geometry of brane Universes see \cite{Ma01}.

Due to the correction terms coming from the extra dimensions,
significant deviations from the Einstein theory occur in brane
world models at very high energies \cite{SMS00}. Gravity is
largely modified at the electro-weak scale $1$ TeV. The
cosmological implications of the brane world theories have been
extensively investigated in the physical literature \cite {all}.
Gravitational collapse can also produce high energies, with the
five dimensional effects playing an important role in the
formation of black holes \cite{all1}.

For standard general relativistic spherical compact objects the
exterior space-time is described by the Schwarzschild metric. In
the five dimensional brane world models, the high energy
corrections to the energy density, together with the Weyl stresses
from bulk gravitons, imply that on the brane the exterior metric
of a static star is no longer the Schwarzschild metric
\cite{Da00}. The presence of the Weyl stresses also means that the
matching conditions do not have a unique solution on the brane;
the knowledge of the five-dimensional Weyl tensor is needed as a
minimum condition for uniqueness. Static, spherically symmetric
exterior vacuum solutions of the brane world models have been
proposed first by Dadhich et al. \cite{Da00} and Germani and
Maartens \cite{GeMa01}. The solution obtained in \cite{Da00} has
the mathematical form of the Reissner-Nordstrom solution, in which
a tidal Weyl parameter plays the role of the electric charge of
the general relativistic solution. A second exterior solution,
which also matches a constant density interior, has been derived
in \cite{GeMa01}. Other classes of exact or approximate (using the
multipole ($1/r$) expansion) solutions of vacuum field equations
on the brane have been obtained in \cite{sol}.

The vacuum field equations on the brane have been reduced to a
system of two ordinary differential equations, which describe all
the geometric properties of the vacuum as functions of the dark
pressure and dark radiation terms (the projections of the Weyl
curvature of the bulk, generating non-local brane stresses) in
\cite{Ha03}. Several classes of exact solutions of the vacuum
gravitational field equations on the brane have been derived, and
vacuums with particular symmetries have been investigated by using
Lie group techniques. A homology theorem for the static,
spherically symmetric gravitational field equations in the vacuum
on the brane has also been proven.

It is the purpose of the present paper to extend the approach
initiated in \cite{Ha03} by considering vacuum space-times on the
brane that are related to some more general Lie groups of
transformations, and to investigate their possible physical
relevance for the explanation of the dynamics of galaxies. As a
group of admissible transformations we chose the one-parameter
group of conformal motions. More exactly, we consider spherically
symmetric and static solutions of the gravitational field
equations for which the metric tensor $g_{\mu \nu }$ has the
property $L_{\xi }g_{\mu \nu }=\psi \left( r\right) g_{\mu \nu }$,
where the left-hand side is the Lie derivative of the metric
tensor, describing the gravitational field in vacuum on the brane,
with respect to the vector field $\xi ^{\mu }$, and $\psi $, the
conformal factor, is an arbitrary function of the radial
coordinate $r$. As for the vector field $\xi ^{\mu }$ we assume
that it is {\it non-static}. With these assumptions the
gravitational field equations, describing the static vacuum brane,
can be integrated in Schwarzschild coordinates, and an exact
solution, corresponding to a brane admitting a one-parameter group
of motions can be obtained. The general solution of the field
equations depends on three arbitrary integration constants. The
conformal symmetry also uniquely fixes the mathematical form of
the dark radiation and dark pressure terms, respectively, which
describe the non-local effects induced by the gravitational field
of the bulk.

As a physical application of the conformally symmetric vacuum
brane model we consider the behavior of the angular velocity of a
test particle moving in a stable circular orbit. It turns out that
for this case the tangential velocity can be expressed as a
function of the conformal factor $\psi $ and some constants of
integration only, the velocity being inversely proportional to the
conformal factor $\psi $. For a specific range of the integration
constants the tangential velocity of the test particle tends, in
the limit of large radial distances, to a constant value. This
behavior is specific to the galactic rotation curves, and is
explained usually by invoking the hypothesis of the dark matter.
However, in the present approach the constant velocity in the
large $r$ limit of a test particle moving in the gravitational
field of a galaxy is due to the existence of the non-local effects
from the bulk, transmitted via the non-zero components of the bulk
Weyl tensor, and of the conformally symmetric geometrical
structure of the static gravitational field on the brane. The
existence of the dark radiation term generates an equivalent mass
term, which is linearly increasing with the distance, and is
proportional to the baryonic mass of the galaxy. All the relevant
physical parameters (metric tensor components, dark radiation and
dark pressure) can be obtained as functions of the tangential
velocity, and hence they can be determined observationally.

The present paper is organized as follows. The basic equations
describing the spherically symmetric gravitational field equations
in the vacuum on the brane are derived in Section II. The general
solution of the vacuum brane space-times admitting a one parameter
group of conformal motions, with non-static conformal symmetry, is
obtained in Section III. The behavior of the angular velocity of a
test particle in stable circular motion is considered in Section
IV. We conclude and discuss our results in Section V.

\section{The field equations for a static, spherically symmetric vacuum brane%
}

On the $5$-dimensional space-time (the bulk), with the negative vacuum
energy $\Lambda _{5}$ and brane energy-momentum as source of the
gravitational field, the Einstein field equations are given by
\begin{equation}
G_{IJ}=k_{5}^{2}T_{IJ},\qquad T_{IJ}=-\Lambda _{5}g_{IJ}+\delta (Y)\left[
-\lambda _{b}g_{IJ}+T_{IJ}^{\text{matter}}\right] ,
\end{equation}
with $\lambda _{b}$ the vacuum energy on the brane and $k_{5}^{2}=8\pi G_{5}$%
. In this space-time a brane is a fixed point of the $Z_{2}$ symmetry. In
the following capital Latin indices run in the range $0,...,4$, while Greek
indices take the values $0,...,3$.

Assuming a metric of the form
$ds^{2}=(n_{I}n_{J}+g_{IJ})dx^{I}dx^{J}$, with $n_{I}dx^{I}=d\chi
$ the unit normal to the $\chi =$constant hypersurfaces and
$g_{IJ}$ the induced metric on $\chi =$constant hypersurfaces, the
effective four-dimensional gravitational equations on the brane
(the Gauss equation), take the form \cite{SMS00}:
\begin{equation}
G_{\mu \nu }=-\Lambda g_{\mu \nu }+k_{4}^{2}T_{\mu \nu }+k_{5}^{4}S_{\mu \nu
}-E_{\mu \nu },  \label{Ein}
\end{equation}
where $S_{\mu \nu }$ is the local quadratic energy-momentum correction
\begin{equation}
S_{\mu \nu }=\frac{1}{12}TT_{\mu \nu }-\frac{1}{4}T_{\mu }{}^{\alpha }T_{\nu
\alpha }+\frac{1}{24}g_{\mu \nu }\left( 3T^{\alpha \beta }T_{\alpha \beta
}-T^{2}\right) ,
\end{equation}
and $E_{\mu \nu }$ is the non-local effect from the free bulk gravitational
field, the transmitted projection of the bulk Weyl tensor $C_{IAJB}$, $%
E_{IJ}=C_{IAJB}n^{A}n^{B}$, with the property $E_{IJ}\rightarrow E_{\mu \nu
}\delta _{I}^{\mu }\delta _{J}^{\nu }\quad $as$\quad \chi \rightarrow 0$. We
have also denoted $k_{4}^{2}=8\pi G$, with $G$ the usual four-dimensional
gravitational constant.

The four-dimensional cosmological constant, $\Lambda $, and the
four-dimensional coupling constant, $k_{4}$, are given by $\Lambda
=k_{5}^{2}\left( \Lambda _{5}+k_{5}^{2}\lambda _{b}^{2}/6\right) /2$ and $%
k_{4}^{2}=k_{5}^{4}\lambda _{b}/6$, respectively. In the limit $\lambda
_{b}^{-1}\rightarrow 0$ we recover standard general relativity.

The Einstein equation in the bulk and the Codazzi equation also imply the
conservation of the energy-momentum tensor of the matter on the brane, $%
D_{\nu }T_{\mu }{}^{\nu }=0$, where $D_{\nu }$ denotes the brane covariant
derivative. Moreover, from the contracted Bianchi identities on the brane it
follows that the projected Weyl tensor should obey the constraint $D_{\nu
}E_{\mu }{}^{\nu }=k_{5}^{4}D_{\nu }S_{\mu }{}^{\nu }$.

The symmetry properties of $E_{\mu \nu }$ imply that in general we can
decompose it irreducibly with respect to a chosen $4$-velocity field $u^{\mu
}$ as \cite{Ma01}
\begin{equation}
E_{\mu \nu }=-k^{4}\left[ U\left( u_{\mu }u_{\nu }+\frac{1}{3}h_{\mu \nu
}\right) +P_{\mu \nu }+2Q_{(\mu }u_{\nu )}\right] ,  \label{WT}
\end{equation}
where $k=k_{5}/k_{4}$, $h_{\mu \nu }=g_{\mu \nu }+u_{\mu }u_{\nu }$ projects
orthogonal to $u^{\mu }$, the ''dark radiation'' term $U=-k^{4}E_{\mu \nu
}u^{\mu }u^{\nu }$ is a scalar, $Q_{\mu }=k^{4}h_{\mu }^{\alpha }E_{\alpha
\beta }$ a spatial vector and $P_{\mu \nu }=-k^{4}\left[ h_{(\mu }\text{ }%
^{\alpha }h_{\nu )}\text{ }^{\beta }-\frac{1}{3}h_{\mu \nu }h^{\alpha \beta }%
\right] E_{\alpha \beta }$ a spatial, symmetric and trace-free tensor.

In the case of the vacuum state we have $\rho =p=0$, $T_{\mu \nu }\equiv 0$
and consequently $S_{\mu \nu }=0$. Therefore, by neglecting the effect of
the cosmological constant, the field equations describing a static brane
take the form
\begin{equation}
R_{\mu \nu }=-E_{\mu \nu },
\end{equation}
with the trace $R$ of the Ricci tensor $R_{\mu \nu }$ satisfying
the condition $R=R_{\mu }^{\mu }=E_{\mu }^{\mu }=0$.

In the vacuum case $E_{\mu \nu }$ satisfies the constraint $D_{\nu
}E_{\mu }{}^{\nu }=0$. In an inertial frame at any point on the
brane we have $u^{\mu }=\delta _{0}^{\mu }$ and $h_{\mu \nu
}=$diag$\left( 0,1,1,1\right) $. In a static vacuum $Q_{\mu }=0$
and the constraint for $E_{\mu \nu }$ takes the form \cite{GeMa01}
\begin{equation}
\frac{1}{3}D_{\mu }U+\frac{4}{3}UA_{\mu }+D^{\nu }P_{\mu \nu }+A^{\nu
}P_{\mu \nu }=0,
\end{equation}
where $D_{\mu }$ is the projection (orthogonal to $u^{\mu }$) of the
covariant derivative and $A_{\mu }=u^{\nu }D_{\nu }u_{\mu }$ is the
4-acceleration. In the static spherically symmetric case we may choose $%
A_{\mu }=A(r)r_{\mu }$ and $P_{\mu \nu }=P(r)\left( r_{\mu }r_{\nu }-\frac{1%
}{3}h_{\mu \nu }\right) $, where $A(r)$ and $P(r)$ (the ''dark
pressure'') are some scalar functions of the radial distance $r$,
and $r_{\mu }$ is a unit radial vector \cite{Da00}.

We chose the static spherically symmetric metric on the brane in the form
\begin{equation}
ds^{2}=-e^{\nu \left( r\right) }dt^{2}+e^{\lambda \left( r\right)
}dr^{2}+r^{2}\left( d\theta ^{2}+\sin ^{2}\theta d\phi ^{2}\right) .
\label{line}
\end{equation}

Then the gravitational field equations and the effective
energy-momentum tensor conservation equation in the vacuum take
the form \cite{Ha03}
\begin{equation}
-e^{-\lambda }\left( \frac{1}{r^{2}}-\frac{\lambda ^{\prime }}{r}\right) +%
\frac{1}{r^{2}}=\frac{48\pi G}{k^{4}\lambda _{b}}U,  \label{f1}
\end{equation}
\begin{equation}
e^{-\lambda }\left( \frac{\nu ^{\prime }}{r}+\frac{1}{r^{2}}\right) -\frac{1%
}{r^{2}}=\frac{16\pi G}{k^{4}\lambda _{b}}\left( U+2P\right) ,  \label{f2}
\end{equation}
\begin{equation}
e^{-\lambda }\left( \nu ^{\prime \prime }+\frac{\nu ^{\prime 2}}{2}+\frac{%
\nu ^{\prime }-\lambda ^{\prime }}{r}-\frac{\nu ^{\prime }\lambda ^{\prime }%
}{2}\right) =\frac{32\pi G}{k^{4}\lambda _{b}}\left( U-P\right) ,  \label{f3}
\end{equation}
\begin{equation}
\nu ^{\prime }=-\frac{U^{\prime }+2P^{\prime }}{2U+P}-\frac{6P}{r\left(
2U+P\right) }.  \label{f4}
\end{equation}

In the following we shall denote $\alpha =16\pi G/k^{4}\lambda _{b}$.

\section{General solution of the vacuum brane field equations with nonstatic
conformal symmetry}

The system of the field equations for the vacuum on the brane is
under-determined. A functional relation between the dark energy
$U$ and the dark pressure $P$ must be specified in order to solve
the equations. An alternative method, which avoids {\it ad hoc }
specifications, is to assume that the brane is mapped conformally
onto itself along the direction $\xi $, so that
\begin{equation}\label{1}
L_{\xi }g_{\mu \nu }=g_{\mu \nu ,\lambda }\xi ^{\lambda
}+g_{\lambda \nu }\xi _{,\mu }^{\lambda }+g_{\mu \lambda }\xi
_{,\nu }^{\lambda }=\psi g_{\mu \nu },
\end{equation}
where $\psi $ is the conformal factor. As for the choice of $\xi
$, Herrera et al. \cite{He} assumed that
\begin{equation}\label{2}
{\bf \xi }=\xi ^{0}\left( r\right) \frac{\partial }{\partial
t}+\xi ^{1}\left( r\right) \frac{\partial }{\partial r}.
\end{equation}

Using this form of the conformal vector in Eqs. (\ref{1}) one
obtains $\xi ^{0}=A$, $\xi ^{1}=\left( B/2\right) r\exp \left(
-\lambda /2\right) $, $\psi \left( r\right) =B\exp \left( -\lambda
/2\right) $ and $\exp \left( \nu \right) =C^{2}r^{2}$, where $A$,
$B$, $C$ are constants. $A$ may be set to zero since $A\partial
/\partial t$ is a Killing vector and $B$ may be set to $1$ by a rescaling $%
\xi \rightarrow B^{-1}\xi $, $\psi \rightarrow B^{-1}\psi $, which leaves
Eqs. (\ref{1}) invariant. This form of $\xi $ gives the most general $\xi $
invariant under the Killing symmetries, that is $\left[ \partial /\partial t,%
{\bf \xi }\right] =0=\left[ {\bf X}_{\alpha },{\bf \xi }\right] $, where $%
{\bf X}_{\alpha }$ generates $SO\left( 3\right) $. This form of
the metric, obtained by imposing static conformal symmetry, has
been used in \cite{MaD} to investigate the properties of strange
stars. The general solution of the vacuum brane gravitational
field equations for this choice of ${\bf \xi } $ has been obtained
in \cite{Ha03}.

A more general conformal symmetry has been proposed by Maartens and
Maharajah \cite{MaMa90}, which generalizes the isotropic conformal vector $%
t\partial /\partial t+r\partial /\partial r$ of the Minkowski
space-time, but weakens the static symmetry of $\xi $ in Eq.
(\ref{2}):
\begin{equation}
{\bf \xi }=\xi ^{0}\left( t,r\right) \frac{\partial }{\partial t}+\xi
^{1}\left( t,r\right) \frac{\partial }{\partial r}.  \label{3}
\end{equation}

Moreover, we assume that the conformal factor $\psi $ is static,
$\psi =\psi \left( r\right) $. With this form of ${\bf \xi }$ Eqs.
(\ref{1}) give immediately (we denote $'=d/dr$):
\begin{equation}\label{41}
\nu ^{\prime }\xi ^{1}+2\frac{\partial \xi ^{0}}{\partial t}=\psi ,
\end{equation}
\begin{equation}
\lambda^{\prime }\xi ^{1}+2\frac{\partial \xi ^{1}}{\partial r}=\psi ,
\end{equation}
\begin{equation}\label{43}
\xi ^{1}=\frac{r\psi }{2}.
\end{equation}

By solving Eqs. (\ref{41})-(\ref{43}) we obtain \cite{MaMa90}
\begin{equation}
\xi ^{0}=A+\frac{1}{2}\frac{k}{B}t,
\end{equation}
\begin{equation}
\psi =Be^{-\lambda /2},
\end{equation}
\begin{equation}\label{nu}
e^{\nu }=C^{2}r^{2}\exp \left( -2kB^{-1}\int \frac{dr}{r\psi
}\right) ,
\end{equation}
where $k$ is a separation constant and $A$, $B$ and $C$ are
integration constants. Without any loss of generality we can chose
$A=0$. Thus for the vector field $\mathbf{\xi }$ we obtain
\begin{equation}
\mathbf{\xi =}\frac{1}{2}\frac{k}{B}t\frac{\partial }{\partial
t}+\frac{r\psi \left( r\right) }{2}\frac{\partial }{\partial r},
\end{equation}
while the metric tensor components of the static vacuum brane can
be expressed as a function of the conformal factor in the form
$\exp \left( \lambda \right) =B^{2}\psi ^{-2}$ and $\exp \left(
\nu \right) =C^{2}r^{2}\exp \left( -2kB^{-1}\int dr/r\psi \right)
$, respectively.

Substitution of these forms of the metric functions in the field equations (%
\ref{f1})-(\ref{f3}) gives
\begin{equation}
-\frac{\psi ^{2}}{B^{2}}\left(
\frac{1}{r^{2}}+\frac{2}{r}\frac{\psi ^{\prime }}{\psi }\right)
+\frac{1}{r^{2}}=3\alpha U,  \label{g1}
\end{equation}
\begin{equation}
\frac{\psi ^{2}}{B^{2}}\left(
\frac{3}{r^{2}}-2\frac{k}{B}\frac{1}{r^{2}\psi }\right)
-\frac{1}{r^{2}}=\alpha \left( U+2P\right) ,  \label{g2}
\end{equation}
\begin{equation}
\psi ^{2}\left( 2\frac{\psi ^{\prime }}{r\psi }-2\frac{k}{B}\frac{1}{%
r^{2}\psi }+\frac{k^{2}}{B^{2}}\frac{1}{r^{2}\psi ^{2}}+\frac{1}{r^{2}}%
\right) =\alpha \left( U-P\right) .  \label{g3}
\end{equation}

By multiplying Eq. (\ref{g3}) by $2$, adding the equation thus
obtained to Eq. (\ref{g2}) and equating the resulting equation
with Eq. (\ref{g1}) gives the following differential equation
satisfied by the function $\psi $:
\begin{equation}
3r\psi \psi ^{\prime }+3\psi ^{2}-3\frac{k}{B}\psi +\frac{k^{2}}{B^{2}}%
-B^{2}=0.  \label{eqf}
\end{equation}
For $k\neq \pm B^{2}$ the general solution of Eq. (\ref{eqf}) is
given by
\begin{equation}\label{r}
r^{2}=R_{0}^{2}\frac{F\left( \psi \right) }{\left| 3\psi ^{2}-3\frac{k}{B}%
\psi +\frac{k^{2}}{B^{2}}-B^{2}\right| },
\end{equation}
where $R_{0}$ is an arbitrary constant of integration,
\begin{equation}
F\left( \psi \right) =\exp \left( -3\frac{k}{B}\int \frac{d\psi
}{3\psi ^{2}-3\frac{k}{B}\psi +\frac{k^{2}}{B^{2}}-B^{2}}\right) ,
\end{equation}
and
\begin{equation}\label{F1}
F\left( \psi \right) =\left( \frac{\left| \psi -\psi _{2}\right|
}{\left| \psi -\psi _{1}\right| }\right) ^{m},k\in \left(
-2B^{2},2B^{2}\right) ,
\end{equation}
\begin{equation}
F\left( \psi \right) =\exp \left( \frac{\pm 2B}{\psi \mp B}\right)
,k=\pm 2B^{2},
\end{equation}
\begin{equation}\label{F3}
F\left( \psi \right) =\exp \left[ -\frac{k}{B}n\arctan n\left( \psi -\frac{k%
}{2B}\right) \right] ,k\in \left( -\infty ,-2B^{2}\right) \cup
\left( 2B^{2},+\infty \right).
\end{equation}

In Eqs. (\ref{F1})-(\ref{F3}) we have also denoted
\begin{equation}
\psi _{1,2}=\frac{3\frac{k}{B}\pm \sqrt{12B^{2}-3\frac{k^{2}}{B^{2}}}}{6},m=%
\frac{3k}{B\sqrt{12B^{2}-3\frac{k^{2}}{B^{2}}}},n=\frac{6}{\sqrt{3\frac{k^{2}%
}{B^{2}}-12B^{2}}}.
\end{equation}

For the dark radiation and dark pressure we obtain the general
expressions
\begin{equation}
U\left( \psi \right) =\frac{\left| 3\psi ^{2}-3\frac{k}{B}\psi +\frac{k^{2}}{%
B^{2}}-B^{2}\right| \left( \psi ^{2}-2\frac{k}{B}\psi +\frac{2}{3}\frac{k^{2}%
}{B^{2}}+\frac{1}{3}B^{2}\right) }{3\alpha B^{2}R_{0}^{2}F\left(
\psi \right) },
\end{equation}
and
\begin{equation}
P\left( \psi \right) =\frac{\left| 3\psi ^{2}-3\frac{k}{B}\psi +\frac{k^{2}}{%
B^{2}}-B^{2}\right| \left( 4\psi ^{2}-2\frac{k}{B}\psi
-\frac{1}{3}\left( \frac{k^{2}}{B^{2}}-B^{2}\right) -2\right)
}{3\alpha R_{0}^{2}F\left( \psi \right) },
\end{equation}
respectively.

Generally, $\psi $ cannot be expressed in an exact analytical form
as a function of $r$. Hence the functions $\exp \left( \lambda
\right) =B^{2}\psi ^{-2}$,
\begin{equation}
\exp \left( \nu \right) =\frac{C^{2}R_{0}^{2}}{F\left( \psi
\right) \left( 3\psi ^{2}-3\frac{k}{B}\psi
+\frac{k^{2}}{B^{2}}-B^{2}\right) },
\end{equation}
$U\left( \psi \right) $ and $P\left( \psi \right) $ can be
obtained, as functions of the radial distance $r$, only in a
parametric form, with $\psi $ taken as parameter.

However, because of the arbitrariness in the choice of the
reference system in the general theory of relativity we can
subject the coordinates to any transformation which does not
violate the central symmetry of the line element. Therefore, by
introducing a new radial coordinate $\bar{r}=\psi \left( r\right)
$, so that
\begin{equation}
r=r\left( \bar{r}\right) =\frac{R_{0}\sqrt{F\left( \bar{r}\right) }}{\sqrt{%
\left|
3\bar{r}^{2}-3\frac{k}{B}\bar{r}+\frac{k^{2}}{B^{2}}-B^{2}\right|
}},
\end{equation}
we obtain the line element of the static, spherically symmetric
metric admitting a conformal symmetry with a non-static vector
field on the vacuum brane in the form
\begin{equation}
ds^{2}=\frac{R_{0}^{2}}{F\left( \bar{r}\right) \left| 3\bar{r}^{2}-3\frac{k}{%
B}\bar{r}+\frac{k^{2}}{B^{2}}-B^{2}\right| }\left[ -C^{2}dt^{2}+\frac{%
9B^{2}F^{2}\left( \bar{r}\right) d\bar{r}^{2}}{\left| 3\bar{r}^{2}-3\frac{k}{%
B}\bar{r}+\frac{k^{2}}{B^{2}}-B^{2}\right| ^{2}}+F^{2}\left(
\bar{r}\right) d\Omega ^{2}\right] ,k\neq \pm B^{2},
\end{equation}
where $d\Omega ^{2}=d\theta ^{2}+\sin ^{2}\theta d\phi ^{2}$ is
the metric of a unit sphere.

Therefore, by using the new variable $\bar{r}$ the three classes
of conformally symmetric solutions of the gravitational field
equations on the brane take the form
\begin{equation}
ds^{2}=\frac{R_{0}^{2}}{3}\frac{\left| \bar{r}-\psi _{1}\right| ^{m-1}}{%
\left| \bar{r}-\psi _{2}\right| ^{m+1}}\left[
-C^{2}dt^{2}+B^{2}\frac{\left|
\bar{r}-\psi _{2}\right| ^{2m-2}}{\left| \bar{r}-\psi _{1}\right| ^{2m+2}}d%
\bar{r}^{2}+\frac{\left| \bar{r}-\psi _{2}\right| ^{2m}}{\left|
\bar{r}-\psi _{1}\right| ^{2m}}d\Omega ^{2}\right] ,k\in \left(
-2B^{2},2B^{2}\right) ,
\end{equation}
\begin{equation}
ds^{2}=\frac{R_{0}^{2}}{3}\frac{\exp \left( \frac{\mp 2B}{\bar{r}\mp B}%
\right) }{\left( \bar{r}\mp B\right) ^{2}}\left[ -C^{2}dt^{2}+B^{2}\frac{%
\exp \left( \frac{\pm 4B}{\bar{r}\mp B}\right) }{\left( \bar{r}\mp
B\right)
^{4}}d\bar{r}^{2}+\exp \left( \frac{\pm 4B}{\bar{r}\mp B}\right) d\Omega ^{2}%
\right] ,k=\pm 2B^{2},
\end{equation}
\begin{eqnarray}
ds^{2} &=&\frac{R_{0}^{2}\exp \left[ \frac{k}{B}n\arctan n\left( \bar{r}-%
\frac{k}{2B}\right) \right] }{\left| 3\bar{r}^{2}-3\frac{k}{B}\bar{r}+\frac{%
k^{2}}{B^{2}}-B^{2}\right| }\times  \nonumber\\
&&\left\{ -C^{2}dt^{2}+\frac{9B^{2}\exp \left[
-2\frac{k}{B}n\arctan n\left(
\bar{r}-\frac{k}{2B}\right) \right] d\bar{r}^{2}}{\left| 3\bar{r}^{2}-3\frac{%
k}{B}\bar{r}+\frac{k^{2}}{B^{2}}-B^{2}\right| ^{2}}+\exp \left[ -2\frac{k}{B}%
n\arctan n\left( \bar{r}-\frac{k}{2B}\right) \right] d\Omega
^{2}\right\} ,\nonumber\\
&& k \in \left( -\infty ,-2B^{2}\right) \cup \left( 2B^{2},+\infty
\right).
\end{eqnarray}

The general solution of the field equations can be obtained in an
exact analytical form for some particular values of $k$. Hence by
taking $k=\pm B^{2}$ we immediately obtain
\begin{equation}
\psi =\frac{R_{0}}{r}\pm B,
\end{equation}
with the corresponding line element given by
\begin{equation}
ds^{2}=\frac{1}{\left( \frac{R_{0}}{r}\pm B\right) ^{2}}\left(
-C^{2}dt^{2}+B^{2}dr^{2}\right) +r^{2}\left( d\theta ^{2}+\sin
^{2}\theta d\phi ^{2}\right) ,k=\pm B^{2}.
\end{equation}

For the dark radiation and the dark pressure we find
\begin{equation}
U\left( r\right) =\frac{1}{\alpha R_{0}^{2}}\left(
\frac{R_{0}}{r}\right) ^{4},P\left( r\right) =\frac{2}{\alpha
R_{0}^{2}}\left( \frac{2R_{0}}{3r}\pm B\right) \left(
\frac{R_{0}}{r}\right) ^{3},k=\pm B^{2}.
\end{equation}

\section{Stable circular orbits in conformally symmetric space-times on
the brane}

We shall consider now the problem of constructing stable circular
timelike geodesic orbits in a static spherically, spherically
symmetric field on the brane, with line element given in a general
form by Eq. (\ref{line}). The motion of a test particle in the
gravitational field can be described by the Lagrangian \cite{Ma03}
\begin{equation}
2L=\left( \frac{ds}{d\tau }\right) ^{2}=-e^{\nu \left( r\right)
}\left(
\frac{dt}{d\tau }\right) ^{2}+e^{\lambda \left( r\right) }\left( \frac{dr}{%
d\tau }\right) ^{2}+r^{2}\left( \frac{d\Omega }{d\tau }\right)
^{2}, \label{lag}
\end{equation}
where we denoted by $\tau $ the affine parameter along the
geodesics. In the timelike case $\tau $ corresponds to the proper
time. In the following we denote by an overdot the differentiation
with respect to $\tau $. From the
Lagrangian given by Eq. (\ref{lag}) it follows that the energy $E=e^{\nu }\dot{t}$ and the $%
\varphi $-component $l_{\varphi }=r^{2}\sin ^{2}\theta
\dot{\varphi}$ of the
angular momentum of the particle are conserved quantities, $E=$const. and $%
l_{\varphi }=$const. The $\theta $-component of the angular momentum, $%
l_{\theta }=r^{2}\dot{\theta}$ is not a constant of the motion,
but the total angular momentum $l^{2}=l_{\theta }^{2}+\left(
l_{\varphi }/\sin \theta \right) ^{2}$ is a conserved quantity,
$l^{2}=$const. The total
angular momentum can be expressed in terms of the solid angle as $l^{2}=r^{4}%
\dot{\Omega}^{2}$ \cite{Ma03}.

In the timelike case the equation of the geodesic orbits can be
written in the form
\begin{equation}
\dot{r}^{2}+V\left( r\right) =0,
\end{equation}
where the potential $V\left( r\right) $ is given by
\begin{equation}
V\left( r\right) =-e^{-\lambda }\left( E^{2}e^{-\nu }-\frac{l^{2}}{r^{2}}%
-1\right) .
\end{equation}

Restricting the radial motion to stable circular orbits implies
imposing the conditions $\dot{r}=0$ and $\partial V/\partial r=0$,
so that the potential describes an extremum of the motion. In
order this extremum be a minimum the condition $\partial
^{2}V/\partial r^{2}>0$ is also required. These three conditions
imply that the circular motion is stable. They also lead to the
following expressions of the energy and total angular momentum of
the particle \cite{Ma03}, \cite{La03}:
\begin{equation}
E^{2}=\frac{2e^{\nu }}{2-r\nu ^{\prime }},l^{2}=\frac{r^{3}\nu ^{\prime }}{%
2-r\nu ^{\prime }}.  \label{cons}
\end{equation}

On the other hand, the line element, given by Eq. (\ref{line}),
can be rewritten in terms of the spatial components of the
velocity, normalized with the speed
of light, measured by an inertial observer far from the source, as $%
ds^{2}=-dt^{2}\left( 1-v^{2}\right) $ \cite{Ma03}, where
\begin{equation}
v^{2}=e^{-\nu }\left[ e^{\lambda }\left( \frac{dr}{dt}\right)
^{2}+r^{2}\left( \frac{d\Omega }{dt}\right) ^{2}\right] .
\end{equation}

For a stable circular orbit $\dot{r}=0$, and the tangential
velocity of the test particle can be expressed as
\begin{equation}
v_{tg}^{2}=\frac{r^{2}}{e^{\nu }}\left( \frac{d\Omega }{dt}\right)
^{2}.
\end{equation}

In terms of the conserved quantities the angular velocity is given
by
\begin{equation}
v_{tg}^{2}=\frac{e^{\nu }}{r^{2}}\frac{l^{2}}{E^{2}}.
\end{equation}

With the use of Eqs. (\ref{cons}) we obtain
\begin{equation}
v_{tg}^{2}=\frac{r\nu ^{\prime }}{2}.
\end{equation}

Thus, the rotational velocity of the test body is determined by
the metric coefficient $\exp \left( \nu \right) $ only.

In the case of the motion of a test particle in a conformally
symmetric, static spherically symmetric space-time, with a
non-static vector field generating the symmetry, the metric
coefficient $\exp \left( \nu \right) $ is given by Eq. (\ref{nu}).
Therefore for the angular velocity we find the simple expression
\begin{equation}
v_{tg}^{2}=1-\frac{k}{B}\frac{1}{\psi }.  \label{tg}
\end{equation}

Eq. (\ref{tg}) gives a simple physical interpretation of the
conformal
factor $\psi $ in terms of the tangential velocity, $\psi =\left( k/B\right) %
\left[\left( 1-v_{tg}^{2}\right)^{-1} \right] $. On the other
hand, the metric coefficient $\exp \left( \lambda \right) $ can
also be expressed as a function of the tangential velocity only:
\begin{equation}
\exp\left( \lambda \right) =\frac{B^{4}}{k^{2}} \left(
1-v_{tg}^{2}\right) ^{2}.
\end{equation}

From Eq. (\ref{tg}) it
follows that the general, physically acceptable, range of the parameter $%
\psi $ is $\psi \in \lbrack k/B,\infty )$, corresponding to a
variation of the tangential velocity between zero and the speed of
light. However, in the case $k\in \left( -2B^{2},2B^{2}\right) $,
the limiting value of the radial coordinate, $r\rightarrow \infty
$, is obtained, as one can see from Eq. (\ref{F1}), in the limits
$\psi \rightarrow \psi _{1}$ or $\psi \rightarrow \psi _{2}$ (the
corresponding limit depends on the numerical values of the
parameters $k$ and $B$). Assuming that $r\rightarrow \infty $ for
$\psi \rightarrow \psi _{1}$, it follows that in the large $r$
limit the tangential velocity of a test particle in stable
circular motion in a conformally symmetric static vacuum
space-time on the brane tends to a limiting, non-zero value
$v_{tg\infty }$, $v\rightarrow v_{tg\infty }$, $r\rightarrow
\infty$, given by
\begin{equation}\label{inf}
v_{tg\infty }=\sqrt{1-\frac{6k}{3k+\sqrt{12B^{4}-3k^{2}}}}.
\end{equation}

For $B=1.00000034$ and $k=0.9999999$ the limiting tangential
velocity is given by $v_{tg\infty }\sim 0.00072\sim 216.3$ km/s,
which is of the order of the observed galactic rotational
velocities.

In the case of a conformally symmetric static vacuum space-time on
the brane, the general dependence of the tangential velocity
$v_{tg}$ on the radial coordinate $r$  is given, with the use of
Eq. (\ref{r}), in a parametric form, with $\psi $ taken as
parameter. In this model it is not
possible to express the tangential  velocity as an analytical function of $r$%
.

The variation of $v_{tg}$ as a function of the radial distance is
represented, for some particular values of $k$ and $B$, in Fig. 1.

\vspace{0.2in}
\begin{figure}[h]
\includegraphics{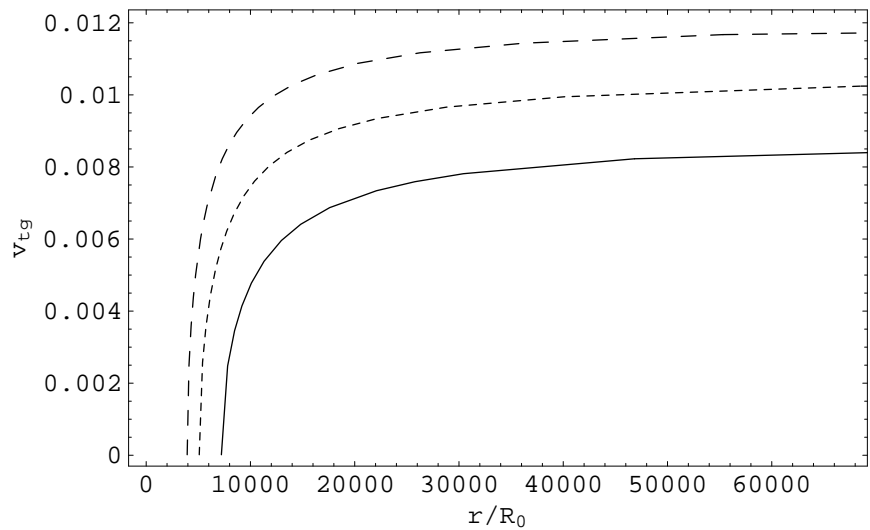}
\caption{Variation, as a function of the parameter $r/R_0$, of the
tangential velocity $v_{tg}$ of a test particle in a stable
circular orbit in a conformally symmetric vacuum space-time on the
brane, for $B=1.00001$ and different values of $k$: $k=0.9999$
(solid curve), $k=0.99985$ (dotted curve) and $k=0.9998$ (dashed
curve). } \label{FIG1}
\end{figure}

In the limit of large $r$, $r\rightarrow \infty$, and for this
choice of the numerical values of the arbitrary parameters $k$ and
$B$, the tangential velocity tends to a constant value. The
numerical value of the limiting velocity is extremely sensitive to
the values of $k$ and $B$.

The variations of the metric coefficients exp$\left(\nu \right)$
and exp$\left(\lambda \right)$ are represented in Figs. 2 and 3,
respectively.

\vspace{0.2in}
\begin{figure}[h]
\includegraphics{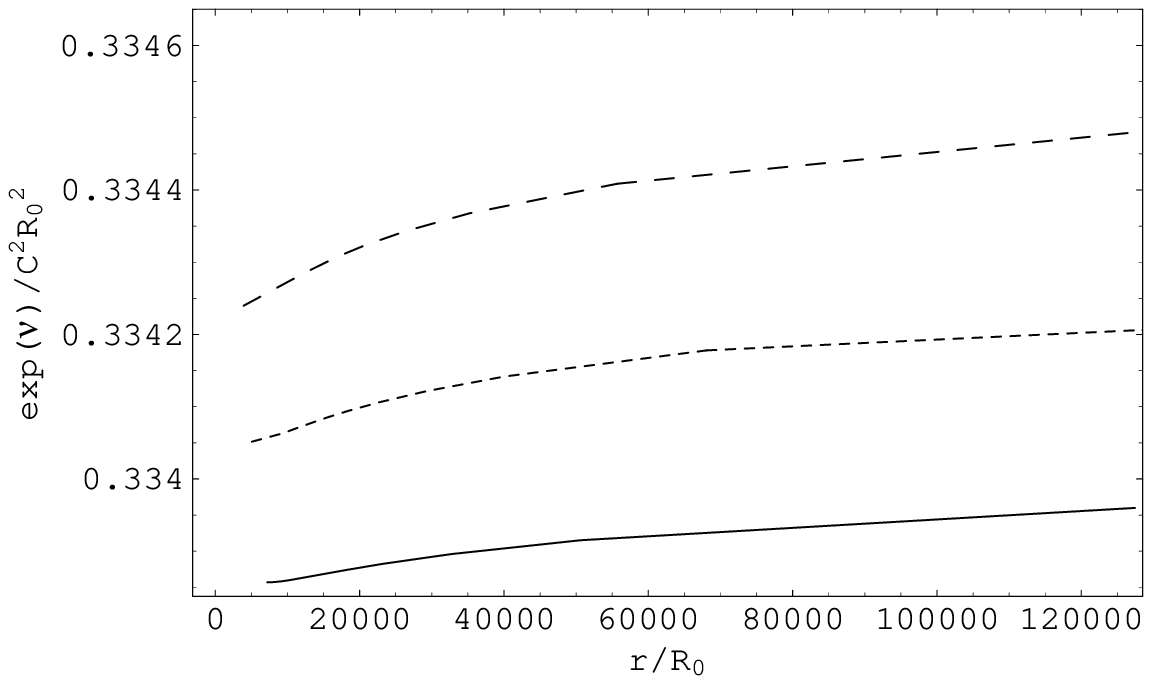}
\caption{Variation, as a function of the parameter $r/R_0$, of the
metric coefficient exp$\left(\nu \right)/C^2R_0^2$ for a static,
conformally symmetric vacuum space-time on the brane, for
$B=1.00001$ and different values of $k$: $k=0.9999$ (solid curve),
$k=0.99985$ (dotted curve) and $k=0.9998$ (dashed curve). }
\label{FIG2}
\end{figure}

\vspace{0.2in}
\begin{figure}[h]
\includegraphics{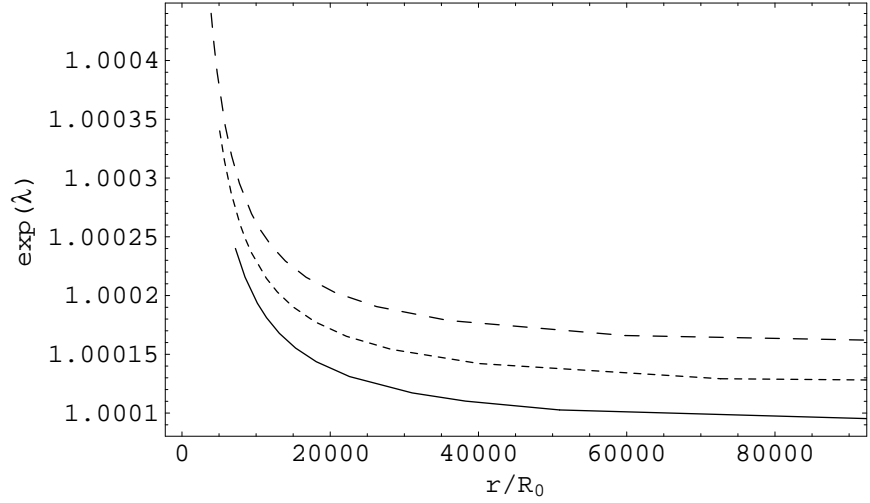}
\caption{Variation, as a function of the parameter $r/R_0$, of the
metric coefficient exp$\lambda $ for a static, conformally
symmetric vacuum space-time on the brane, for $B=1.00001$ and
different values of $k$: $k=0.9999$ (solid curve), $k=0.99985$
(dotted curve) and $k=0.9998$ (dashed curve). } \label{FIG3}
\end{figure}

The metric components satisfy the conditions exp$\left(\nu
 \right)\leq 1$ and exp$\left(\lambda \right)\geq 1$, respectively.

The dark radiation and dark pressure terms can also be
represented, as a function of the tangential velocity of a test
body, in the form
\begin{equation}
U\left( v_{tg}\right) =\frac{\left\{ B^{4}\left( 1-v_{tg}^{2}\right) +k^{2}%
\left[ 2v_{tg}^{2}\left( 1+v_{tg}^{2}\right) -1\right] \right\} \left| k^{2}%
\left[ 1+v_{tg}^{2}\left( 1+v_{tg}^{2}\right) \right] -B^{4}\left(
1-v_{tg}^{2}\right) ^{2}\right| }{9\alpha R_{0}^{2}B^{6}\left(
1-v_{tg}^{2}\right) ^{4}F\left[ \frac{k}{B}\left( 1-v_{tg}^{2}\right) ^{-1}%
\right] },
\end{equation}
and
\begin{equation}
P\left( v_{tg}\right) =\frac{\left\{ B^{2}\left( B^{2}-6\right)
\left(
1+v_{tg}^{2}\right) +k^{2}\left[ 5+v_{tg}^{2}\left( 8-v_{tg}^{2}\right) %
\right] \right\} \left| k^{2}\left[ 1+v_{tg}^{2}\left( 1+v_{tg}^{2}\right) %
\right] -B^{4}\left( 1-v_{tg}^{2}\right) ^{2}\right| }{9\alpha
R_{0}^{2}B^{4}F\left[ \frac{k}{B}\left( 1-v_{tg}^{2}\right)
^{-1}\right] },
\end{equation}
respectively.

The variation of the dark radiation $U$ is represented, as a
function of $r/R_0$, in Fig. 4.

\vspace{0.2in}
\begin{figure}[h]
\includegraphics{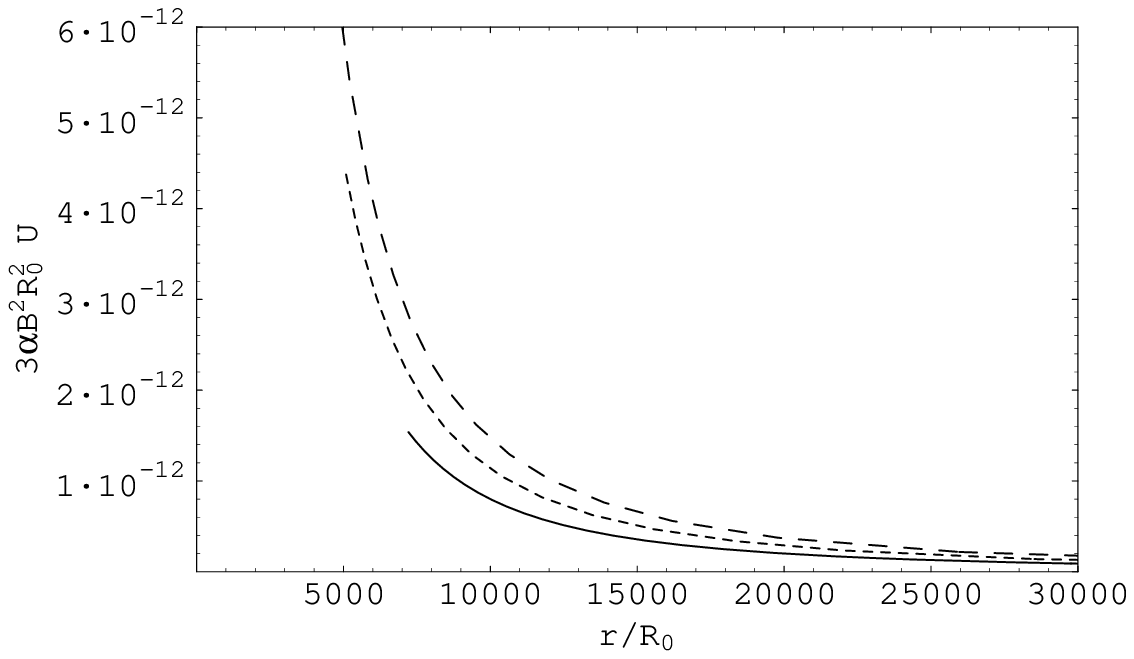}
\caption{Variation, as a function of the parameter $r/R_0$, of the
dark radiation term $3\alpha B^2R_0^2U$ for a static, conformally
symmetric vacuum space-time on the brane, for $B=1.00001$ and
different values of $k$: $k=0.9999$ (solid curve), $k=0.99985$
(dotted curve) and $k=0.9998$ (dashed curve). } \label{FIG4}
\end{figure}

The dark radiation term is positive for all values of the radial
coordinate $r$, $U(r)\geq 0$, $\forall r\in \left( 0,\infty
\right) $. In the limit of large $r$, $U$ tends to zero,
$\lim_{r\rightarrow \infty }U(r)=0$. The variation of the dark
pressure as a function of $r$ is represented in Fig. 5.

\vspace{0.2in}
\begin{figure}[h]
\includegraphics{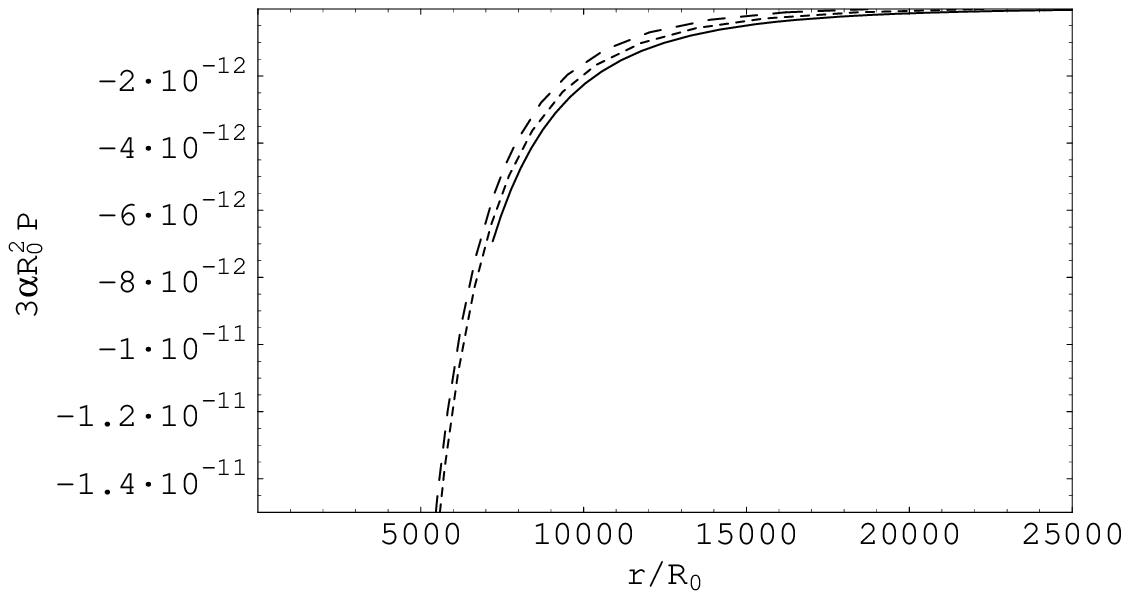}
\caption{Variation, as a function of the parameter $r/R_0$, of the
dark pressure term $3\alpha R_0^2P$ for a static, conformally
symmetric vacuum space-time on the brane, for $B=1.00001$ and
different values of $k$: $k=0.9999$ (solid curve), $k=0.99985$
(dotted curve) and $k=0.9998$ (dashed curve). } \label{FIG5}
\end{figure}

In the present model the dark pressure is negative, satisfying the
condition $P(r)\leq 0$, $\forall r\in \left( 0,\infty \right) $.
In the large time limit, similar to the dark radiation term, the
dark pressure also tends to zero, $\lim_{r\rightarrow \infty
}P(r)=0$.

\section{Discussions and final remarks}

In order to obtain a manifestly coordinate invariant
characterization of certain geometrical properties of geometries,
like for example curvature singularities, Petrov type of the Weyl
tensor etc., the scalar invariants of the Riemann tensor have been
extensively used. Two scalars, which have been
considered in the physical literature, are the Kretschmann scalars, $%
RiemSq\equiv R_{ijkl}R^{ijkl}$ and $RicciSq\equiv R_{ij}R^{ij}$, where $%
R_{ijkl}$ is the Riemann curvature tensor.

For space-times which are the product of two 2-dimensional spaces,
one Lorentzian and one Riemannian, subject to a separability
condition on the function which couples the 2-spaces, it has been
suggested in \cite{Sa98} that the set $C=\left\{
R,r_{1},r_{2},w_{2}\right\}$ form an independent set of scalar
polynomial invariants, satisfying the number of degrees of freedom
in the curvature. $R=g^{il}g^{jk}R_{ijkl}$ is the Ricci scalar and the quantities $r_{1}$, $%
r_{2}$ and $w_{2}$ are defined according to \cite{Za01}:
\begin{equation}
r_{1}=\phi _{AB\dot{A}\dot{B}}\phi ^{AB\dot{A}\dot{B}}=\frac{1}{4}%
S_{a}^{b}S_{b}^{a}, r_{2}=\phi _{AB\dot{A}\dot{B}}\phi _{C\dot{C}}^{B\dot{B}}\phi ^{CA\dot{C}%
\dot{A}}=-\frac{1}{8}S_{a}^{b}S_{b}^{c}S_{c}^{a},
\end{equation}
\begin{equation}
w_{2}=\Psi _{ABCD}\Psi _{EF}^{CD}\Psi ^{EFAB}=\frac{1}{32}\left(
3E_{b}^{a}H_{c}^{b}H_{a}^{c}-E_{b}^{a}E_{c}^{b}E_{a}^{c}\right) +\frac{i}{32}%
\left(
H_{b}^{a}H_{c}^{b}H_{a}^{c}-3E_{b}^{a}E_{c}^{b}H_{a}^{c}\right),
\end{equation}
where $S_{a}^{b}=R_{a}^{b}-\frac{1}{4}R\delta _{a}^{b}$ is the
trace-free
Ricci tensor, $\phi _{AB\dot{A}\dot{B}}$ denotes the spinor equivalent of $%
S_{ab}$, $\Psi _{ABCD}$ denotes the spinor equivalent of the Weyl tensor $%
C_{abcd}$ and $\bar{C}_{abcd}$ denotes the complex conjugate of
the self-dual Weyl tensor, $C_{abcd}^{+}=\frac{1}{2}\left(
C_{abcd}-i\ast C_{abcd}\right) $. $E_{ac}=C_{abcd}u^{b}u^{d}$ and
$H_{ac}=C_{abcd}^{\ast }u^{b}u^{d}$ are the "electric" and
"magnetic" parts of the Weyl tensor, respectively, where $u^{a}$
is a timelike unit vector and $C_{abcd}^{\ast }=\frac{1}{2}\eta
_{abef}C_{cd}^{ef}$ is the dual tensor. The expressions of the
invariants for some particular values of the integration constants
$k$ and $B$ are presented in the Appendix. Due to their
complicated form we shall not present the values of the invariants
for other values of $k$ and $B$. For the $B=1$, $k=2$ case the
invariants diverge at $r=1$, while for the $B=1$, $k=\pm 1$ case
they diverge for $r\rightarrow 0$.

In the present paper we have obtained all the conformally
symmetric solutions of the vacuum field equations in the brane
world model, under the assumption of a non-static conformal
symmetry, and we have discussed some of their physical properties.
In particular we have considered the behavior of the angular
velocity of a test particle in stable circular orbit on the brane.
The conformal factor $\psi $, together with two constants of
integration, uniquely determines the rotational velocity of the
particle. In the limit of large radial distances and for a
particular set of values of the integration constants the angular
velocity tends to a constant value. This behavior is typical for
massive particles (hydrogen clouds) outside galaxies. Thus the
rotational galactic curves can be naturally explained in brane
world models. The galaxy is embedded in a modified, spherically
symmetric geometry, generated by the non-zero contribution of the
Weyl tensor from the bulk. The extra-terms, which can be described
in terms of a dark radiation term $U$ and a dark pressure term
$P$, act as a "matter" distribution outside the galaxy. The
particles moving in this geometry feel the gravitational effects
of $U$ and $P$, which can also be described, equivalently, by
means of a mass term.

The behavior of the metric coefficients and of the angular
velocity in the solutions we have obtained depend on two arbitrary
constants of integration $k$ and $B$. Their numerical value can be
obtained by assuming the continuity of the metric coefficient
$\exp \left( \lambda \right) $ across the vacuum boundary of the
galaxy. For simplicity we assume that inside the "normal"
(baryonic) luminous matter, with density $\rho _B$, which form a
galaxy, the non-local effects of the Weyl tensor can be neglected.
We define the vacuum boundary $r_0$ of the galaxy (which for
simplicity is assumed to have spherical symmetry) by the condition
$\rho _B \left(r_0\right)\approx 0$. Therefore at the vacuum
boundary the metric coefficient $\exp \left( \lambda
\right)=1-2GM_B/r_0$, where $M_{B}=4\pi \int_{0}^{r_{0}}\rho
_{B}\left( r\right) r^{2}dr$ is the total baryonic mass inside the
radius $r_0$. The continuity of $\exp \left( \lambda \right) $
through the surface $r=r_0$ gives
\begin{equation}
1-\frac{2GM_{B}}{r_{0}}=\frac{\psi ^{2}\left( r_{0}\right) }{B^{2}}=\frac{%
k^{2}}{B^{4}}\frac{1}{\left[ 1-v_{tg}^{2}\left( r_{0}\right)
\right] ^{2}},
\end{equation}
leading to
\begin{equation}\label{kB}
\frac{k^{2}}{B^{4}}=\left( 1-\frac{2GM_{B}}{r_{0}}\right) \left[
1-v_{tg}^{2}\left( r_{0}\right) \right] ^{2}.
\end{equation}

Therefore the ratio $k^{2}/B^{4}$ can be determined
observationally. With the help of Eq. (\ref{kB}) the limiting
angular velocity of the test particle rotating in the conformally
symmetric gravitational field on the brane, given by Eq.
(\ref{inf}), can be expressed, as a function of the total baryonic
mass of the galaxy only, in the form
\begin{equation}
v_{tg\infty }=\sqrt{1-\frac{6\sqrt{1-\frac{2GM_{B}}{r_{0}}}\left( 1-\frac{%
GM_{B}}{r_{0}}\right) }{3\sqrt{1-\frac{2GM_{B}}{r_{0}}}\left( 1-\frac{GM_{B}%
}{r_{0}}\right) +\sqrt{12-3\left( 1-\frac{2GM_{B}}{r_{0}}\right) \left( 1-%
\frac{GM_{B}}{r_{0}}\right) ^{2}}}},
\end{equation}
where we have also used the Newtonian approximation
$v_{tg}^{2}\left( r_{0}\right) =GM_{B}/r_{0}$ to eliminate the
angular velocity of a test particle at the vacuum boundary of the
galaxy.

Since for a galaxy $GM_{B}/r_{0}$ has a very small value,  we can expand $%
v_{tg\infty }$ in a power series of $GM_{B}/r_{0}$, thus obtaining
\begin{equation}
v_{tg\infty }\approx \frac{2}{\sqrt{3}}\sqrt{\frac{GM_{B}}{r_{0}}}+\frac{1}{%
12\sqrt{3}}\left( \frac{GM_{B}}{r_{0}}\right) ^{3/2}+O\left[ \left( \frac{%
GM_{B}}{r_{0}}\right) ^{5/2}\right] .  \label{vin}
\end{equation}

For a galaxy with baryonic mass of the order $10^{9}M_{\odot }$
and radius of the order of $r_{0}\approx 70$ kpc, Eq. (\ref{vin})
gives $v_{tg\infty }\approx 287$ km/s, which is of the same order
of magnitude as the observed value of the angular velocity of the
galactic rotation curves.

From the field equation Eq. (\ref{f1}) it follows that in the
vacuum outside the galaxy the metric tensor component $\exp \left(
-\lambda \right) $ can be expressed in terms of the dark radiation
only as $\exp \left( -\lambda \right) =1-2GM_{U}/r$, where
$M_{U}=3\alpha \int_{r_{0}}^{r}U(r)r^{2}dr$ represents the "mass"
associated to the dark radiation component of the energy-momentum
tensor on the brane. By using the conformal symmetry and the
expression of the ratio $k^{2}/B^{4}$ we obtain for $M_{U}$ the
expression
\begin{equation}
M_{U}(r)=\frac{r}{2G}\left\{ 1-\left(
1-\frac{2GM_{B}}{r_{0}}\right) \left[ \frac{1-v_{tg}^{2}\left(
r_{0}\right) }{1-v_{tg}^{2}\left( r\right) }\right] ^{2}\right\}.
\end{equation}

Since $v_{tg}^{2}\left( r_{0}\right) $ and $v_{tg}^{2}\left(
r\right) $ are much smaller than one,  it follows that the dark
radiation mass can be approximated by the very simple scaling
relation
\begin{equation}
M_{U}(r)\approx M_{B}\frac{r}{r_{0}}.
\end{equation}

$M_{U}$ is linearly increasing with the distance and is
proportional to the baryonic mass of the galaxy. In the Newtonian
limit, from the equality between the centrifugal force and the
gravitational force it follows $M_{B}/r_{0}=v_{tg}^{2}\left(
r_{0}\right) /G$, leading to
\begin{equation}
M_{U}\left( r\right) \approx \frac{v_{tg}^{2}\left( r_{0}\right)
}{G}r.
\end{equation}

In conclusion, in the present paper we have investigated
conformally symmetric vacuum solutions of the gravitational field
equations on the brane, and analyzed the motion of test particles
in stable circular orbits in this geometry. By using the
continuity of the metric coefficients a complete description of
the motion of the particles outside a galaxy can be obtained. In
the large distance limit the angular velocity of the particles
tend to a constant value, which can be determined as a function of
the baryonic (luminous) mass and the radius of the galaxy. All the
relevant physical quantities, including the dark energy and the
dark pressure terms, which describe the non-local effects due to
the gravitational field of the bulk, are expressed in terms of
observable parameters. More general conformally symmetric
solutions on the brane, and their physical properties, will be
considered in detail in a future publication.

\section*{Appendix}

In this Appendix we present the values of the Kretschmann scalars $%
RiemSq\equiv R_{ijkl}R^{ijkl}$ and $RicciSq\equiv R_{ij}R^{ij}$and
some values of the independent set of the scalar polynomial
invariants $\left\{ R,r_{1},r_{2},w_{2}\right\} $ for the exact
static, spherically symmetric vacuum brane with conformal symmetry
for some particular values of the integration constants $k$ and
$B$.

For $k=2$ and $B=1$ the expressions of the invariants are
\begin{equation}
R=0,RicciSq=18R_{0}^{-4}e^{-\frac{4}{r-1}}\left( r-1\right)
^{4}\left( 9-8r+12r^{2}-16r^{3}+6r^{4}\right) ,
\end{equation}
\begin{equation}
RiemSq=72R_{0}^{-4}e^{-\frac{4}{r-1}}\left( r-1\right) ^{4}\left(
6-4r+6r^{2}-8r^{3}+3r^{4}\right) ,
\end{equation}
\begin{equation}
r_{1}=\frac{1}{4}RicciSq,
r_{2}=\frac{81}{4R_{0}^{6}}e^{-\frac{6}{r-1}}\left( r-1\right)
^{6}\left( r^{2}-2\right) \left( 1-4r+2r^{2}\right) ,
\end{equation}
\begin{equation}
\text{Re}\left( w_{2}\right) =-%
\frac{81}{4R_{0}^{6}}e^{-\frac{6}{r-1}}\left( r-1\right) ^{6}.
\end{equation}

For $B=1$ and $k=\pm 1$ we obtain
\begin{equation}
R=0,RicciSq=\frac{4R_{0}^{2}\left( 6r^{2}\pm 8R_{0}r+3R_{0}^{2}\right) }{%
r^{8}},
\end{equation}
\begin{equation}
RiemSq=2RicciSq,
\end{equation}
\begin{equation}
r_{1}=\frac{RicciSq}{4},r_{2}=\frac{3R_{0}^{3}\left( 2r^{3}\pm
5R_{0}r^{2}+4R_{0}^{2}r\pm R_{0}^{3}\right) }{r^{12}},
\end{equation}
\begin{equation}
\text{Re}\left( w_{2}\right) =0.
\end{equation}

\end{document}